\documentclass[aps,prl,twocolumn]{revtex4}
\usepackage{graphicx}
\begin{document}

\title{Full counting statistics of electron transfer between
  superconductors}

\author{W. Belzig$^{1,2}$ and Yu.~V. Nazarov$^1$}
\address{$^1$Department of Applied Physics and Delft Institute of
Microelectronics and Submicrontechnology,\\ Delft University of
Technology, Lorentzweg 1, 2628 CJ Delft, The Netherlands\\
$^2$ Department of Physics and Astronomy, University of Basel,
Klingelbergstr. 82, 4056 Basel, Switzerland}

\begin{abstract}
We present an extension of the Keldysh-Green's function method, which
allows to calculate the full distribution of transmitted particles
through a mesoscopic superconductor. The method is applied to the
statistics of supercurrent in short contacts. If the current is
carried by Andreev bound states the distribution corresponds to
switching between long trains of electrons going in opposite
directions. For weak (gapless) superconductors or tunnel junctions we
find that at low temperatures the distribution has negative
``probabilities''. Accounting for the quantum mechanical nature of the
measuring device shows that these negative values can indeed
be measured.\\[2mm]
\hspace*{35mm}{\small [Published as Phys. Rev. Lett. {\bf 87}, 197006 (2001)]}
\end{abstract}


\maketitle

Coherent charge transfer between superconductors (S), supercurrent, is
essentially a quantum-mechanical process. Although superconducting
junctions are commonly used, the statistical properties of the charge
transfer involved into the supercurrent are not yet completely
understood. In view of recent attempts to use the coherence of
superconductors to build quantum bits\cite{qubits}, it is necessary to
reveal the basic limitations on this coherence (if there are any).
Additionally, the problem of the statistics of transferred charge in a
quantum process is of fundamental interest. It is related to the
understanding of the measurement process and the interpretation of its
outcome.

Recently the current noise exhibited in SNS junctions, where N is a
diffusive normal metal, was addressed experimentally.\cite{strunk} The
experimental results show a giant excess noise in the low temperature
and voltage regime in those samples, in which at the same temperature a
coherent coupling through the normal metal was measured. This is in
accordance with theoretical predictions for the shot noise in short
contacts.\cite{scsnoise} This may hint to the importance of an
understanding of the statistical properties of the supercurrent in such
junctions. The equilibrium noise properties have been studied in
\cite{supernoise} and \cite{averin:96}.  Further experimental progress
in the fabrication of controllable single-channel junctions is to be
expected in the near future. This will shed more light on the
fundamental statistical properties of charge transfer between
superconductors.

We will make use of the so called {\em full counting statistics} (FCS),
originally introduced to calculate the distribution of transmitted
charge through a contact between normal metals\cite{levitov:96}. This
theory allows to find the cumulant generating function (CGF) $S(\chi)$,
from which the distribution of transmitted charge follows via $P(N)=\int
d\chi \exp(-S(\chi)-iN\chi)$.  It is tempting (and has been done so far)
to interpret $P(N)$ as the probability that $N$ charges have been
transferred through the contact during the time of observation. We will
show below that this interpretation is strict only for normal
constrictions. For superconducting constrictions the distribution also 
depends on the phase difference $\phi$. It turns out that $P(N,\phi)$ can
also take negative values, which hampers such interpretation. This
is related to the fact that the phase and number of charges transferred
can be regarded as canonically conjugated variables. Still $P(N,\phi)$
provides a complete description of all charge transfer processes and can
be extracted from the results of measurements.

The most powerful and general method of calculating transport properties
of mesoscopic conductors is the nonequilibrium Green's function approach
(see \cite{rammer}). It was shown in \cite{yuli:99-2} that this approach
can be generalized to access FCS. In this Letter we extend the approach
to superconductors.  This allows us to obtain the FCS of an arbitrary
mesoscopic conductor at all temperatures and voltages. The CGF is
derived for a contact which is fully characterized by an ensemble of
transmission eigenvalues $\{T_n\}$. We evaluate the FCS of supercurrent
in two generic cases. First, we find the distribution of transmitted
charge of a single channel contact between two gapped superconductors.
Here the current is carried by phase dependent Andreev bound states and,
as shown by our analysis, conforms with the switching
picture.\cite{averin:96} The two bound states carrying current in
opposite directions are alternately occupied and charges are transfered
in 'long trains', which reflects the coherent nature of the
supercurrent. In the second case we calculate the CGF of a contact
between two weak superconductors. The resulting CGF corresponds to the
tunnel limit for gapped superconductors and can be related to the
effective Keldysh action of a Josephson junction discussed in detail in
Ref.~\cite{schoenzaikin}.  The standard interpretation\cite{levitov:96}
of the CGF leads in the low temperature regime to negative
``probabilities'' $P(N,\phi)$. Negative values of $P(N,\phi)$ occur
because of an attempt to interpret the quantum mechanical phenomenon of
supercurrent with classical means.  If we account for the quantum
mechanical nature of the measuring device, we can resolve the paradox
and specify how $P(N,\phi)$ can be measured.

\begin{figure}[htbp]
  \begin{center}
    \includegraphics[width=6cm]{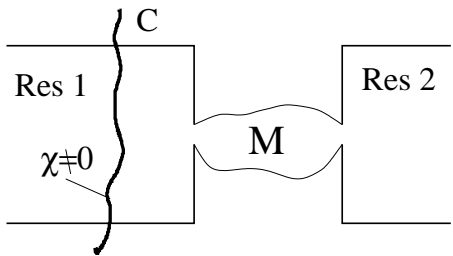}
    \caption{Sketch of the system. Two reservoirs (1,2) are connected to a
      mesoscopic conductor M. The counting field $\chi$ is chosen
      nonzero on the cross section C in reservoir 1.}
    \label{fig:system}
  \end{center}
\end{figure}

To be specific let us now introduce our model system, which is depicted
in Fig.~\ref{fig:system}. A mesoscopic conductor is placed between two
reservoirs.  The counting field $\chi$ is introduced on an arbitrary
cross section in one of the reservoirs and couples to the operator of
current through that cross section. It follows from the definition of the
cumulants that the CGF can be found from
\begin{equation}
  e^{-S(\chi,\phi)}=\left\langle 
    {\cal T} e^{i\frac{\chi}{2}\int_0^{t_0}\hat I(t)dt}
    \tilde{\cal T} e^{i\frac{\chi}{2}\int_0^{t_0}\hat I(t)dt}
  \right\rangle\,.
\end{equation}
Here ${\cal T}( \tilde{\cal T})$ denotes the (anti)time ordering
operator.  $\hat I$ denotes the current operator $\int d^3x
\hat\Psi^\dagger \bar\tau_3 (\bbox{p}/m)\hat \Psi^\dagger
\bbox{\nabla} F(\bbox{x})$, where $\hat\Psi$ is the usual Nambu spinor
field operator and $\bar\tau_3$ is a matrix in Nambu space.
$\bbox{\nabla}F$ is chosen such, that the spatial integration
is restricted to the cross section and yields the total current. The
counting field parameterized in this way can now be incorporated into
the boundary condition imposed by the reservoir onto the mesoscopic
conductor.\cite{yuli:99-2} That is, the reservoir Green's function
effectively takes the form
\begin{equation}
  \label{eq:bc1}
  \check G_1(\chi,\phi) = 
  e^{\frac{i}{2}\chi\check\tau_{\rm K}}
  \check G_1(\phi)
  e^{-\frac{i}{2}\chi\check\tau_{\rm K}}\;.
\end{equation}
Here $\check G_1(\phi)$ is the reservoir Green's function at
superconducting phase $\phi$ in the absence of the counting field and
$\check\tau_{\rm K}=\hat\sigma_3\bar\tau_3$ a matrix in
Keldysh($\hat{\ }$)-Nambu($\bar{\ }$) space. Now the counting field is
included into the boundary condition for the Keldysh-Nambu matrix
Green's functions provided by the left reservoir.  Inside the system
of interest the transport properties are described by quasiclassical
Eilenberger equations \cite{eilenberger}, applicable if the system size
exceeds the Fermi wave length. It is important that $\check
G(\chi,\phi)$ still obeys the quasiclassical normalization condition
$\check G^2(\chi,\phi)=\check 1$.

In certain cases the action $S(\chi,\phi)$ can be found quite generally.
One example is a constriction shorter than the coherence length, which
is fully characterized by a set of transmission eigenvalues $\{T_n\}$.
The counting field manipulates the matrix structure of the Green's
functions in Keldysh-Nambu space. To find the transport properties one
should therefore use expressions, which respect the full matrix
structure. It was noted in \cite{yuli:99}, that a convenient way to do
this is to use a ``matrix current'', which is conserved in short
contacts. The matrix current is formed with the current operator and the
corresponding matrix elements of the Green's functions.  Physical
currents are related to certain components of the matrix current.  For
our purpose here, we need the matrix current in a short contact.
The matrix current was derived in \cite{yuli:99} in the absence of the
counting field. The counting field does not change the matrix structure
of that result. So we can use it just by including the
$\chi$-dependence of the Green's functions and write
\begin{equation}
  \label{eq:matrix-current}
  \check{I}(\chi,\phi)=\frac{1}{2\pi}\sum_n\int dE 
  \frac{T_n\left[\check G_1(\chi,\phi),\check G_2\right]}{
    4+T_n\left(\{\check G_1(\chi,\phi),\check G_2\}-2\right)}\,.
\end{equation}
The action can then be found from the relation
$(\partial/\partial\chi)S(\chi,\phi) = -it_0 \text{Tr}(\check{\tau}_{\rm
  K}\check I(\chi,\phi))$. Using the fact that $[\check A,\{\check
A,\check G_2\}]=0$ for all matrices with $\check A^2=\check 1$, it is
easy to verify that under the trace in (\ref{eq:matrix-current})
$(\partial/\partial\chi) \{\check G_1(\chi,\phi),\check G_2\} =
i\check\tau_K\left[\check G_1(\chi,\phi),\check G_2\right]$. We can
therefore integrate Eq.~(\ref{eq:matrix-current}) with respect to $\chi$
and obtain
\begin{equation}
  \label{eq:action}
  S(\chi,\phi)=\frac{-t_0}{2\pi}\sum_n\int dE \text{Tr}
  \ln\left[4+T_n\left(\{\check G_1(\chi,\phi),\check G_2\}-2\right)\right]\;.
\end{equation}
Eq.~(\ref{eq:action}) is very general. It contains the statistical
properties of all types of superconducting constrictions.  For instance
the FCS of an SN-contact \cite{khmelnitzkii} can easily be obtained from
(\ref{eq:action}).

In the rest of the paper we will study equilibrium noise and statistics of
systems with two superconducting contacts. We will distinguish two generic
cases. The first will be a single-mode contact of arbitrary transparency
between two fully gapped superconductors. In the second case we treat a
contact between two weak superconductors, or, which is equivalent, a tunnel
contact. The channel summation in the action (\ref{eq:action}) is then a
trivial summation over transparencies. In the following derivation we limit
the discussion to a single channel of transmission $T_1$ and identical
reservoirs at equilibrium.  To be specific, we consider the Green's functions
of the reservoirs
\begin{equation}
  \label{eq:reservoir}
  \check G_S= \frac{\bar R+\bar A}{2}
  +\frac{\bar A-\bar R}{2}
  \left(
    \begin{array}[c]{cc}
      - h & (1-h)\\
      (1+h)& h
    \end{array}\right).
\end{equation}
Here $\bar R(\bar A)(E)$ are retarded and advanced Green's functions of
the banks and $h(E)=\tanh(E/2T)$ accounts for the equilibrium
distribution at a temperature $T$. The phase difference $\phi$ is
introduced by setting $\check G_1(\phi) = \exp(i\phi\bar\tau_3/2) \check G_S
\exp(-i\phi\bar\tau_3/2)$ and $\check G_2=\check G_{\text{S}}$.
Advanced and retarded functions in (\ref{eq:reservoir}) possess the
structure $\bar R(\bar A) = g_{\text{R,A}}\bar\tau_3 +
f_{\text{R,A}}\bar\tau_1$ fulfilling the normalization condition
$f^2+g^2=1$. They depend on energy and the superconducting order parameter
$\Delta$. Their precise forms will be defined below.

The trace in the action can be evaluated and we obtain the main result
of this paper
\begin{equation}
  \label{eq:action2}
  S(\chi,\phi)=\frac{-t_0}{\pi}\int dE 
  \ln\left[1+\sum_{n=-2}^2 \frac{A_n(\phi)}{Q(\phi)}
    \left(e^{in\chi}-1\right)\right] \,.
\end{equation}
Introducing $q=(1-g_{\rm R}g_{\rm A})(1-h^2) +f_{\rm R}f_{\rm A}(1+h^2)$
the coefficients may be written as
\begin{eqnarray}
  \label{eq:coeffa2}
  A_{\pm 2} & = & \frac{T^2_1}{64}q^2,\\
  \label{eq:coeffa1}
  A_{\pm 1} & = &  \frac{T_1}{4}q-
  \frac{T^2_1}{16}q\left[q-4f_{\rm R}f_{\rm A}\sin^2\frac\phi2\right]
  \\\nonumber&&
  +\frac{T_1}{8}\left[(f_{\rm R}+f_{\rm A})h\cos\frac\phi2\mp 
    i(f_R-f_A)\sin\frac\phi2\right]^2,\\
  \label{eq:coeffq}
  Q & = & \left[1-T_1f_{\rm R}^2\sin^2\left(\frac{\phi}{2}\right) \right]
  \left[1-T_1f_{\rm A}^2\sin^2\left(\frac{\phi}{2}\right)\right].
\end{eqnarray}
The interpretation of the different terms is analogous to that given in
\cite{khmelnitzkii}. A coefficient $A_{\pm n}$ is related to events in
which a charge $n$ is transfered to the right(left). The presence of
terms, which describe charge transfers of 2e, is a consequence of
superconducting correlations. The interpretation of these terms as
probabilities stems from the comparison with the case of binomial
statistics (see \cite{levitov:96}). As we will discuss below this
interpretation only makes sense for normal metals.

Considering only the phase independent terms of
(\ref{eq:action2})-(\ref{eq:coeffq}) demonstrates an interesting feature.
These terms can be factorized into
\begin{equation}
  \label{eq:normal}
  \left[1+\frac{T_1q}{8}\left(e^{i\chi}-1\right)+
    \frac{T_1q}{8}\left(e^{-i\chi}-1\right)\right]^2\,.
\end{equation}
Due to the logarithm in Eq.~(\ref{eq:action2}), the exponent of 2 can be
written as prefactor to the CGF. Therefore, the CGF describes two
statistically independent probabilistic processes, which we can identify with
electron and hole transfers.  The phase dependent terms inhibit this
factorization and lead to correlations between electrons and holes, as
expected in superconductors. The denominator $Q$, common to all coefficients,
has roots for energies, at which Andreev bound states exist. Consequently the
statistical properties are dominated by the charge transfer through Andreev
bound states. The corresponding ``probabilities'' can be very large, in
particular, larger than 1. Thus, we can not interpret the coefficients
(\ref{eq:coeffa2})-(\ref{eq:coeffq}) as probabilities anymore. To find the
statistics of the charge transfer, we have to specify the system further.

{\em Gapped superconductors.} If the two leads are gaped like BCS
superconductors the spectral function are given by $f_{\text{R,A}} =
i\Delta/((E\pm i\delta)^2-\Delta^2)^{1/2}$ and $g_{\text{R,A}}$ follows from
normalization. Here $\delta$ is an broadening parameter, which accounts for
the finite lifetime of the Andreev bound states due to {\em e.~g.} phonon
scattering. The supercurrent is solely carried by Andreev bound states with
energies $\pm\Delta(1-T_1\sin^2\phi/2)^{1/2} \equiv \pm E_{\text{B}}(\phi)$.
The importance of these bound states can be seen from the coefficient $Q$
(\ref{eq:coeffq}).  It may become zero and will thus produce singularities in
the action.\cite{bs} The broadening $\delta$ shifts the singularities of $Q$
into the complex plane and allows an expansion of the coefficients in
Eq.~(\ref{eq:coeffa2})-(\ref{eq:coeffq}) close to that energy.  Performing the
energy integration the action results in
\begin{equation}
  \label{eq:bs-action1}
  S(\chi,\phi)=-2t_0\delta\sqrt{1-I_1^2(\phi)\chi^2/4\delta^2
    -i\chi \bar I_1(\phi)/\delta}\,,
\end{equation}
where $I_1(\phi)=\Delta^2 T_1 \sin(\phi)/2E_{\text{B}}(\phi)$ is the
supercurrent carried by one bound state and $\bar I_1(\phi) =
I_1(\phi) \tanh(E_{\text{B}}/2T)$ is the average current through the
contact. In deriving (\ref{eq:bs-action1}) we have also assumed that
$\chi\ll 1$. This corresponds to a restriction to ``long trains'' of
electrons transfered, and the discreteness of the electron transfer
plays no role here.  Fast switching events become less probable at low
temperatures and are neglected here.  In the saddle point
approximation at low temperatures $\gamma\equiv1/\cosh(E_B/2T)\ll1$ we
find for the current distribution
\begin{equation}
  \label{eq:super-distr}
  P(j,\phi)\sim\frac{1}{\gamma}
  e^{2\delta t_0\left(\gamma\sqrt{1-j^2(\phi)}-j(\phi)\sqrt{1-\gamma^2}\right)}
  \,,
\end{equation}
for $|j(\phi)|\le 1$ and zero otherwise.
Here, we have expressed the transfered charge in terms of the current
normalized to the zero temperature supercurrent: $j(\phi)=I/I_1(\phi)$. The
current is related to the particle number by $N=It_0$. At zero temperature
Eq.~(\ref{eq:super-distr}) approaches $P(j,\phi)\to\delta(j-1)$, which follows
from a direct calculation.  Thus, at zero temperature the current is noiseless
and the distribution (\ref{eq:super-distr}) at finite temperature confirms the
picture of switching between Andreev states which carry current in opposite
directions, suggested in Ref.~\cite{averin:96}.

Let us finally comment on the limits under which the previous result is
valid. In the energy integration it was assumed that the bound states
are well defined. For small transmission the distance of the bound state
to the gap edge is $\approx T_1\Delta$. Thus, to have well defined bound
states we have to require $\delta < T_1\Delta$. Similarly for a high
transmissive contact and a small the phase difference we require
$\phi\sim I/I_{\rm c} > \delta/\Delta$. The statistics beyond these
limits is similar to what is discussed in the following.

{\em Tunnel junction/gapless superconductors.} Let us now consider the
supercurrent statistics between two weak superconductors, where the Green's
functions can be expanded in $\Delta$ for all energies. One can see that this
is equivalent to the tunneling limit ($\{T_n\}\ll 1$) of
Eq.~(\ref{eq:action}). We also return the many channel situation here.
Expanding the action (\ref{eq:action}) to lowest order and using that the
counting rotation can be written as $ \exp (i\chi\check\tau_K/2) = \frac 12
\left[e^{i\chi/2}(1-\check\tau_K) + e^{-i\chi/2}(1+\check\tau_K)\right]$ we
find
\begin{equation}
  \label{eq:tunnel-action}
    S=-t_0
    \left[iI_{\rm s}(\phi)\sin\chi + 
      P_{\rm s}(\phi)\left(\cos\chi-1\right)\right]\;.
\end{equation}
In short, the full statistics are expressed in terms of supercurrent
$I_{\rm s}(\phi)$ and noise $P_{\rm s}(\phi)$.
In equilibrium using (\ref{eq:reservoir})
\begin{eqnarray}
  \label{eq:tunnel-current}
  I_{\rm s}(\phi) &=&-\frac{G}{4}\textrm{Re}\int dE 
  \text{Tr}\left\{\bar\tau_3\left[\bar R_1(\phi),\bar R_2\right]\right\}h\\
  \label{eq:tunnel-noise}
  P_{\rm s}(\phi) &=&-\frac{G}{4}\textrm{Re}\int dE 
  \text{Tr}\left\{\bar\tau_3\bar A_1(\phi)\bar\tau_3\bar R_2\right\}(1-h^2)
\end{eqnarray}
Here $G=(1/\pi)\sum T_n$ is the normal state conductance of the contact.
The equivalence of this result to the limit of gapless superconductors
follows from an expansion of (\ref{eq:action})-(\ref{eq:coeffq}) to
orders $f^2$.

Eq. (\ref{eq:tunnel-noise}) shows that $P_{\rm s}$ vanishes at zero
temperature, since $h(T=0)=\pm 1$, whereas $I_{\rm s}$ vanishes at $T_{\rm
  c}$.  Therefore, there is some crossover temperature below which $P_{\rm
  s}<I_{\rm s}$. In this limit the action possesses no saddle point anymore
and by expansion in powers of $\exp(i\chi)$ it follows, that $P(N,\phi)$
becomes negative.  Obviously this questions the direct interpretation of
$P(N,\phi)$ as a probability. Thus, we are forced to have a closer look on
what $P(N,\phi)$ actually is.

To clarify this issue, we make use of the recent results presented in
\cite{kindermann}, where it was shown that the interpretation of $P(N,\phi)$
in intimately related to the way the measurement is performed. We assume a
simple model of a measuring device: a capacitor of infinite capacitance that
stores the charge passed through the constriction, i.~e. the charge operator
$\hat q$ is related to the current operator through the constriction by
$\stackrel{.}{\hat q}=\hat I$. The quantum mechanical treatment of this device
involves its density matrix $\rho(q,q')$.

In \cite{kindermann} the relation between initial and final density matrices
of the device was obtained. This can be expressed in terms of the density
matrix in Wigner representation, $\rho(x,q)$, $x$ being the canonical
conjugate of $q$. It reads
\begin{equation}
\rho_f(x,q) = \sum_N P(N,\phi - x) \rho_i(x,q -N)\,,
\label{for_rho}
\end{equation} 
so that $P(N,\phi)$ fully characterizes the quantum mechanical behavior of the
capacitor. For a normal constriction $P(N,\phi)$ does not depend on $\phi$. In
this case we can rewrite Eq.~(\ref{for_rho}) directly in terms of charge
distributions $\Pi(q) \equiv \int dx \rho(x,p)$,
\begin{equation}
\Pi_f(q) = \sum_N P(N) \Pi_i (q-N)\,.
\label{for_pi}
\end{equation}
Therefore, $P(N)$ can be interpreted as {\it classical} probability.  For a
superconducting constriction quantum mechanics is essential and the resulting
charge distribution depends on the details of $\rho_i$.  For instance, if one
sets $\rho(q,q')$ to $\delta(q)\delta(q')$ the probabilities $\Pi_f(q)$ do not
depend on $\phi$:
\begin{equation}
\Pi_f(q)= \sum_N \delta(q-N) \int d \phi P(N,\phi)\,.
\end{equation}
A similar result for a simple Josephson junction model was cited in
\cite{schoenzaikin}.  A more general choice of $\rho_i$ preserves the
$\phi$-dependence.  One can summarize the situation by saying that $N$
and $\phi$ are related to canonically conjugated variables $q$ and
$x$, that hampers their simultaneous measurement.
 
Since $\rho (x,p)$ are not positive in general, the $P(N,\phi)$ do not
have to be positive.  It might seem that these "negative
probabilities" can not be measured.  Fortunately, it is not so.  To
understand this, let us see how one would measure $P(N)$ in the {\it
  classical} case. The only exact way is to make use of
Eq.~(\ref{for_pi}).  One thus measures $\Pi_{i,f}$ separately and then
obtains $P(N)$ from a deconvolution procedure: the Fourier transform
of $P$ is the ratio of Fourier transforms of $\Pi$'s.  Our main result
is that the same deconvolution procedure can be applied to
Eq.~(\ref{for_rho}), resulting in
\begin{equation}
  \label{eq:deconv}
  P(N,\phi)=\int\frac{d \chi}{2\pi} e^{iN\chi}
  \frac{\rho_f(\phi+\chi/2,\phi-\chi/2)}{\rho_i(\phi+\chi/2,\phi-\chi/2)}\,.
\end{equation}
Since off-diagonal entries of the density matrix can not be measured, this
expression is not directly applicable. In \cite{kindermann} a scheme was
proposed, how this can be circumvented by a repeated measurement of
differently prepared initial density matrices. This allows one to characterize
and measure $P(N,\phi)$, whatever sign it has.

In conclusion we have studied the statistical properties of supercurrent
in short constrictions. An extension of the Keldysh technique to account
for full counting statistics of systems containing superconductors was
developed. In the case of the supercurrent through a short constriction
(point contact or tunnel junction) the cumulant generating function can
be found quite generally. It shows that charge transfer occurs in units
of $e$ and $2e$, which largely enhanced probabilities in the case of
contacts with large transmission. The charge transfer occurs in long
trains of electrons passing through the contact in either direction,
which is a signature of the coherent nature of the supercurrent. The
relative probability of trains in the two directions is determined by
the thermal occupation and the switching rate between them by the
broadening parameter in the bulk of the superconductors. For tunnel
junctions or point contacts between gapless superconductors we find the
occurrence of negative values of $P(N,\phi)$, which questions the
interpretation as a probability. Accounting for the full time evolution
of an (idealized) measuring device, we have shown that these negative
values can indeed be observed.

We acknowledge useful discussions with L.~S.~Levitov and
C.~W.~J.~Beenakker.  W.B. was financially supported by the ``Stichting
voor Fundamenteel Onderzoek der Materie'' (FOM) and the ``Alexander von
Humboldt-Stiftung''.


\begin{thebibliography}{99}

\bibitem{qubits}
  A. Shnirman, Z. Hermon, and G. Sch\"on, Phys. Rev. Lett. {\bf 79},
  2371 (1997); J.~E. Mooij {\em et al.}, Science {\bf 285}, 1036 (1999);
  L.~B. Ioffe {\em et al.}, Nature {\bf 398}, 679 (1999).

\bibitem{strunk}
   T. Hoss {\em et al.}, Phys. Rev. B {\bf 62}, 4079 (2000).

\bibitem{scsnoise}
  J.~C. Cuevas, A. Mart\'{\i}n-Rodero, and A. Levy Yeyati,
  Phys. Rev. Lett. {\bf 82}, 4086 (1999); Y. Naveh and D.~V. Averin,
  {\em ibid.}, 4090.

\bibitem{supernoise}
  A. Mart\'{\i}n-Rodero, A. Levy Yeyati, and F.~J. Garc\'{\i}a-Vidal,
  Phys. Rev. B {\bf 53}, R8891 (1996).

\bibitem{averin:96}
  D. Averin and H.~T. Imam, Phys. Rev. Lett. {\bf 76}, 3814 (1996). 

\bibitem{levitov:96} L. S. Levitov, H. W. Lee, and G. B. Lesovik,
  Journ. Math. Phys. {\bf 37}, 4845 (1996).
  
\bibitem{rammer}  
  J. Rammer and H. Smith,
  Rev. Mod. Phys. {\bf 58}, 323 (1986).

\bibitem{yuli:99-2}
  Yu. V. Nazarov, 
  Ann. Phys. (Leipzig) {\bf 8} Spec. Issue, SI-193 (1999).

\bibitem{schoenzaikin}
  G. Sch\"on and A.~D. Zaikin, Phys. Rep. {\bf 198}, 237 (1990).
  
\bibitem{yuli:99} Yu. V.  Nazarov, Superlattices Microst.\ {\bf
    25}, 1221 (1999).

\bibitem{eilenberger}
  G. Eilenberger,
  Z. Phys. {\bf 214}, 195 (1968);
  A.~I. Larkin and Yu.~N. Ovchinnikov,
  Sov. Phys. JETP {\bf 26}, 1200 (1968);  
  K.~D. Usadel, 
  Phys. Rev. Lett. {\bf 25}, 507 (1970).

\bibitem{khmelnitzkii} B. A. Muzykantskii and D. E. Khmelnitzkii,
  Phys. Rev. B {\bf 50}, 3982 (1994). 
  
\bibitem{bs} It is interesting to note that in the limit $\delta\to 0$
  the action has poles for energies $E_B^2(\chi) = \Delta^2 \left(1-T
    \sin^2\left( \frac{\phi\pm\chi}{2} \right)\right)$. The counting
  field therefore couples directly to the phase sensitivity of the
  Andreev bound states.

\bibitem{kindermann}
  Yu. V. Nazarov and M. Kindermann, cond-mat/0107133 (unpublished).
\end{thebibliography}
\end{document}